\documentclass{ws-procs975x65}

\def\lsim{\lower.5ex\hbox{$\; \buildrel < \over \sim \;$}}
\def\gsim{\lower.5ex\hbox{$\; \buildrel > \over \sim \;$}}

\begin{document}

\title{Role of Disk models in Indentifying Astrophysical Black Holes}

\author{S. K. CHAKRABARTI\footnote{\uppercase{A}lso at \uppercase{C}entre for 
\uppercase{S}pace \uppercase{P}hysics, \uppercase{C}halantika 43, \uppercase{G}aria 
\uppercase{S}tation \uppercase{R}d., \uppercase{K}olkata 700084}}

\address{S.N. Bose National Centre for Basic Sciences,\\
JD-Block, Salt Lake, Kolkata 700098, INDIA\\
E-mail: chakraba@bose.res.in\\ }

\maketitle

\abstracts {We discuss how disk models may limit the scope of identifying 
astrophysical black holes. We show that the standard Keplerian thin disk model,
the thick disk model, slim disks, ADAFs etc. are fundamentally limited.
We present the most complete solution to date called the advective accretion 
disk and discuss how it has the scope to address every observational aspects 
of a black hole. Though the magnetic field is not fully self-consistently 
taken care of yet, the details with which the present model 
can handle various issues successfully are astounding. We present some of the examples.}

\noindent To be Published in the proceedings of the 10th Marcel Grossman Meeting (World Scientific Co., Singapore),
Ed. R. Ruffini et al.

\section{Introduction}

Celestial bodies satisfying the classical black hole solutions of Einstein's field equations
are supposed to be the most enigmatic objects in the sky. They exhibit their 
presence only through gravitational attraction. There is no 
spectral information directly from a classical black hole. A non-rotating 
black hole has only one parameter, namely, the mass. A rotating black hole
has two parameters: the mass and the specific angular momentum. Though there 
are exciting discussions on charged black holes (having three parameters, such as 
mass, charge, and specific angular momentum) being the sources of GRBs (see,
recent papers of R. Ruffini and collaborators, this volume), because of complexity, 
the advective disks have not been studied around this type of black holes yet.

In this review paper, we shall discuss a few accretion flow models and show that 
some of them may not be suitable for identifying black holes in the first place. Some models may  
have more ways to explore black hole behaviour than the other. We believe that 
the advective disk models which explain the steady and time dependent spectral 
characteristics and which also produce jets and outflows are the best models
to investigate a black hole property. First, we discuss the basic properties of 
a black hole, and then we discuss the properties of the major disk models 
which are in the literature today. We then discuss the intrinsic limitations
of each of these models.

\section{Properties of a Black Hole And The Nearby Disk Matter}

A classical non-rotating (Schwarzschild) black hole is compact and of size $r_g=2GM/c^2$
and the light crossing time is merely $r_g/c=2GM/c^3=10^{-5}\frac{M}{M_\odot}$s. 
For a Kerr black hole, the light crossing time is even shorter and could be up to half as
much when one has an extremely rotating Kerr black hole. The inflow radial velocity
on the horizon, is the velocity of light and this is independent of the outer boundary 
condition (Chakrabarti 1990a). Half of the photons emitted within the photon orbit 
$r_{ph}=1.5r_g$ of a black hole are `sucked in'. Part of the photons emitted outside 
$r_{ph}$ will also be `sucked in' depending on the impact parameter. 

An important aspect of the accretion onto a black hole is that even when the energy of the flow is 
{\it totally} conserved, the solution is perfectly stable (Chakrabarti, 1989). What this means 
is that the flow need not radiate and can quietly allowed itself inside the hole without any 
observable signature. On a neutron star surface, this is not possible. The flow must `hit' the
hard surface and radiate all of its kinetic energy by generating heat and radiation. Thus, a realistic 
accretion disk model must allow almost constant energy flow. 

Another important aspect, often wrongly questioned, is that a black hole can have a `boundary layer'
even though it does not have a hard surface. This was introduced and established by Chakrabarti
and his collaborators over past fifteen years. If the flow is totally radial, the hot and compressed flow
may produce pairs and the pressure could slow down the inflowing gas, thereby producing a standing or moving shock
(Kazanas and Elison, 1986). The X-rays emitted at the inner-edge can also heat the outer skirts of the
flow and slow it down and produce shocks (Chang and Ostriker, 1985). A third and the most important possibility
occurs when the flow has an angular momentum  which is not dissipated fast enough  (either because of 
low viscosity, or because the infall velocity is too high, i.e., close to a black hole). The centrifugal
force slows down the matter and the matter is `piled up' behind the centrifugal barrier (Chakrabarti 1989, 1990a,
1996a). Fig. 1 shows a cartoon picture of this advective disks.
We shall show below that those models which do not incorporate such a possibility are fundamentally 
flawed and must be rejected. Indeed, all the observations point to the necessity of these two behaviours.

\begin{figure}[t]
\vskip 0.0cm
\centerline{\epsfxsize=5.5in\epsfbox{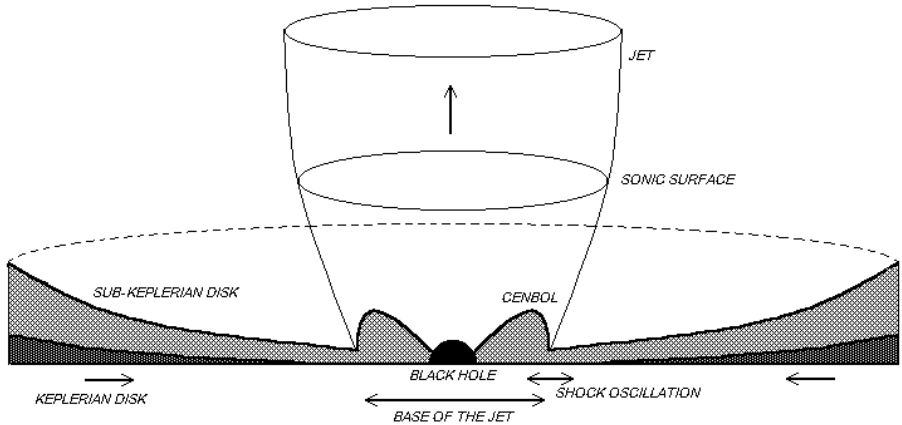}}
\vskip -1.2cm
\caption{Cartoon diagram of the Two Component Advective Flow  (TCAF) solution around a black hole.} 
\end{figure}

\section{Properties of Accretion Flows Around a Black Hole} 

\subsection{Keplerian Disk}

So far, we discussed how the matter should behave close to the horizon. Far away, it is expected that 
the properties of matter should be independent of whether there is a black hole at the centre for not.  
Thus, for instance, a disk with matter distributed in Keplerian orbits (Shakura and Sunyaev, 1973)
will not distinguish between a black hole and an ordinary star even at a moderate distance away from the
central object. Furthermore, since a Keplerian disk is terminated at the marginally stable orbit
($r_{mb}=3r_g$ for a non-rotating black hole) it does not `sample' the horizon. A truly Keplerian disk
is subsonic and  very slowly accreting. Though there are ample evidences that such disks do exist
in galactic and extra-galactic systems which contain black holes, there are more compelling reasons to believe that 
such disks cannot be the whole story. This is partly because the contribution of the Keplerian disk cannot
change in a matter of a few seconds and the rapid variation in the power-law hard tail clearly requires
another component in the flow. 

\subsection{Disks with a Corona?}

The meaning of the word `corona' is vague. For instance one could ask (a) Is it always present? (b) Is it
moving like a flow? (c) In what time-scale does it change? (d) Is this magnetic in origin, just like stars?
In other words, what is the mechanism to produce it? There is no answer, and it is likely there will be none.
This generic term is constantly used in the literature to blissfully push the problem away and to 
cover up the ignorance about the hydrodynamics and magnetohydrodynamics of
the problem. If the corona is of solar type, who is anchoring 
the ends of the corona? The constantly moving flow cannot do this, because the infall time close to the
hole is comparable to the buoyancy time and the whole flux tube would be popped out of the disk.
A disk without a radiative core and a convective envelope (as in the sun) cannot keep a corona in stellar
sense. Thus the so-called disk-corona cannot be of magnetic origin. 
If it made out of hot plasma, is it static? Can a static corona survive on the top of a rapidly 
inflowing disk? Is the plasma coming out of a Keplerian disk because of evaporation? 
If so, what is the rate of evaporation? Is the evaporation because intrinsic 
`boiling' of the disk, or due to irradiation of X-rays emitted further close to the hole, or both?
The observations seem to indicate that the corona is dynamic, it is changing its character 
very fast, in a free-fall time (Smith et al. 2001, 2002). Thus, the only possibility is that 
it is another flow which engulfs the Keplerian disk on the equatorial plane. The interaction
between the Keplerian disk and this `halo', both from the hydrodynamical and from the radiative point of view
should be interesting to study and a lot of work has been done (Chakrabarti and Titarchuk, 1995;
Chakrabarti 1997; Chakrabarti 1998a) in this direction. 

\subsection{Nature of the Halo: A Sub-Keplerian Advective Flow}

The `halo' that Chakrabarti \& Titarchuk (1995) introduced is not arbitrary. If there is going to be another
component, it got to be a sub-Keplerian component, because a super-Keplerian component (a component having
specific angular momentum more than a Keplerian disk everywhere) would be thrown out by the 
centrifugal  force.
A sub-Keplerian component, on the contrary would have lower angular momentum and would be falling
faster than a Keplerian disk. A sub-Keplerian component can be generated in many ways: (a) In a binary system, 
if the companion is losing matter through winds, then the angular momentum  $l$ 
of this wind is on an average `zero' with respect to the companion. 
It becomes a finite but a small number with respect to the compact primary. (b) In the case of active 
galactic nuclei, the stars around will loose winds and the wind will collide to cancel out their 
transverse velocities (Chakrabarti \& Molteni, 1994, unpublished). On an average, only the radial
velocities will remain.  This would also create a sub-Keplerian flow around a massive black hole. 
Unlike in the stellar system, where, some binaries may undergo Roche-lobe overflow and therefore
are guaranteed to have a Keplerian disk, in AGNs, perhaps the sub-Keplerian component is the only 
dominant component around. (c) Even if the initial flow is Keplerian, the viscous mechanism can be
such that it transports more angular momentum outwards and creates a sub-Keplerian flow. The theoretical 
work in this line is simple (Chakrabarti 1990a; 1996a): only for viscosity coefficients above a certain value 
the sub-Keplerian flow can become Keplerian. For lower viscosity coefficients the disk is sub-Keplerian. (d) A sub-Keplerian 
flow (a spherical Bondi flow being a special case) which starts from a large distance
must have a positive specific energy. A sub-Keplerian flow which starts 
close to the hole should have a positive energy. Whereas a cool, Keplerian disk always has negative energy,
a hot Keplerian disk, or a disk away from the equatorial plane can have positive specific energy. Thus,
energetically it is not at all unlikely to generate sub-Keplerian flows on the way to the black hole.
(e) Magnetic dissipation can play an important role in making positive energy, sub-Keplerian 
flow out of a negative energy flow. It is well argued (Chakrabarti 1990b)
that magnetic pressure increases more rapidly than the ion pressure as the flow moves 
towards a black hole. If one assumes that the magnetic energy density cannot be 
more than the equipartition value, then most of the magnetic energy must be dissipated away
and surely a part would heat up the disk. Cumulative effect would be to bring matter 
out of the clutch of a Keplerian disk and free them into the sub-Keplerian halo.
(f) In any case, the flow must be super-sonic on the horizon (Chakrabarti 1990a) and
therefore it must deviate from a Keplerian disk much before entering through the horizon.
This deviation is a must even if the energy is negative.  The actual solution topologies of
all possible solutions are discussed in detail in Chakrabarti (1996a-c).

This sub-Keplerian matter thus originated will flow almost freely till it hits the
centrifugal barrier given by $l^2/r^3 \sim 1/(r-2)^2$ (here we assumed that a Paczy\'nski and Wiita (1980) type
potential is satisfactory enough for a non-rotating hole). This would be strictly valid for 
particles. For fluids, the barrier will be farther away from the hole because of the flow pressure. 
The virial temperature of this sub-Keplerian flow is high enough to create a `moving hot corona' 
sandwiching the Keplerian disk. This component may be called the advective flow, because of the radial
motion or the transonic flow, because it passes through the sonic points one or more times. Together,
the sandwiched disk is called the Two Component Advective Flow, or TCAF. The Keplerian disk is a
special case of an advective disk. The region in between the horizon and the centrifugal barrier 
is called the Centrifugal Pressure Dominated BOundary Layer or simply CENBOL. 

\subsection{Thick Accretion Disks and their connection to Modern Advective Flows}

In the late seventies, Paczy\'nski and their collaborators (see, e.g., Paczy\'nski and Wiita (1980)) 
realized that a flow {\it could} become sub-Keplerian in some region and super-Keplerian in some 
other region. Assuming a {\it pure} rotational motion, they came up with the concept of a thick accretion 
disk. Simply put, the flow is assumed to puff up due to radiation pressure and the disk can be supported 
against falling into a black hole even when the angular momentum is lower than the Keplerian far away.
These disks have a toroidal topology and were fascinating to look at and to draw the contours of 
constant pressure, temperature etc., though it was not clear what to do with them. This was because,
the models studied were non-accreting. They  do have the {\it possibility} of accretion via 
Roche-lobe like open potential surfaces. Secondly, the energy and angular momentum were 
nearly constant. But they had advantages that one could imagine that the ets or outflows
could be produced along the axis of the torus and they could be easily collimated by the torus wall
(so-called funnel wall). For high accretion rates, the Polish models were valid. For lower accretion 
rates, the British model (see, Rees et al. 1982) showed that the ion pressure would be 
much better. However, without a viable accretion or jet mechanism it was piece of a design or sculpture 
to look at. The generalization of thick disk solutions in General Relativistic framework can be seen 
in Chakrabarti (1985).

There were unsuccessful attempts to view these thick disks as a part of the Keplerian disk. What was realized
back in 1993 (Chakrabarti 1993) was that these thick disks are actually the special cases of the 
CENBOL region of an advective flow. In the CENBOL, the flow becomes sub-sonic (thus only
rotational dominated) and hot. 
The disk is puffed up and it takes the shape of a torus with virtually the same contour 
shapes (Molteni, Sponholz and Chakrabarti, 1994; Molteni, Ryu and Chakrabarti, 1996). 
However, unlike a thick disk, which is just rotating, the CENBOL flow
eventually becomes supersonic and enters through the horizon.  Furthermore, as will be discussed below,
this CENBOL is a part of the inflow coming from a large distance. This CENBOL can be cooled down by
photons and the spectral properties depend on its existence. It can oscillate to cause X-ray oscillation
or the so-called quasi-periodic oscillation. It can produce jets and outflows etc. So, though thick
disks HAD some characteristics of the CENBOL, it was incomplete. 

\subsection{Slim Disks}

Much is said about the so-called slim disk (Abramowicz et. al. 1988). It is pertinent to ask why it was
brought in and what it accomplished so far. A 
slim disk is basically the end product of a line of thought
pushed by Paczy\'nski and his collaborators who felt that the Keplerian disk should be extended 
inwards till the horizon. Furthermore, there were apprehensions that the standard Keplerian disk 
could be unstable (Shapiro, Lightman and Eardley, 1976) and one could check if the `realistic' version 
of it, i.e., the one which reaches the horizon may also have the same instability. It was found
that indeed, the inward flow, which advected away the perturbation inside, does stabilize the disk.
This solution was obtained for very (perhaps unrealistically) high accretion rate and the 
disk was otherwise having the characteristics of a Keplerian disk. It is unclear whether these
solutions have any application in any of the black hole candidates known today.

\subsection{Advection Dominated Accretion Flows (ADAFs)}

The discussion becomes incomplete if, along with Polish and British models mentioned above, an American
model is not discussed. A model was proposed which was originally self-similar (Narayan \& Yi, 1994) and 
later the equations were integrated along the line of already established transonic solutions.
Here, the heat generated in the flow is supposed to be advected inside the black hole. 
A contingent of papers flooded the subject, all claiming that 
all the spectra of all the black holes, stellar or extragalactic, could be explained by ADAFs. Today
we know that absolutely nothing is explained by ADAFs. Instead, this rat-race to join these modelists
only delayed the progress of the subject several years. Slowly every `evidence' of ADAF 
became `non-evidence' and every models needed a `patch-up'. ADAFS started 
to change its name to CDAF, JDAF, WADAF etc. Most interestingly, even though ADAFs were proposed,
and indeed were found to be stable for very low accretion rates (Park and Ostriker, 2002), 
several workers with completely different disk models (including those with high accretion rates)
wanted to join the band-wagon and started calling themselves as ADAFists. Similarly, though, originally 
ADAFers thought that shocks should not exist, solutions started coming in from 
various corners (e.g., Yuan, Markoff \& Falcke 2002) that shocks could be allowed and jets are actually shock
processed matter. Thus, the features of the advective flow models of Chakrabarti (1999) 
where it was shown that outflows are produced only from CENBOL, i.e., post-shock regions, are 
already reappearing in the so-called ADAF models. 

Another important issue was the geometry of accretion flow. According 
to ADAF model (Narayan, 1997) the inner region should be spherical and 
the flow should deviate from a Keplerian flow very far away. This is 
clearly wrong, since a funnel wall must form along the axis even when
the matter is of very low angular momentum. After Chakrabarti and 
Titarchuk (1995) and Ebisawa, Titarchuk \& Chakrabarti (1996) spelled
out how a TCAF model should look like geometrically, the above wrong 
designs were quickly corrected and several `cartoon'  pictures of ADAF 
model came out in the literature which are clearly `inspired' by TCAF
(Novak 2003). Thus, it was getting clear that ADAF was converging to 
already existing successful advective flow solution and that it had
nothing `new' to offer.

Blows came from other sides also: Theoreticians found that the truly 
correct ADAF solution could be obtained more easily using the method 
of Chakrabarti \& his collaborators (Lu, Gu \& Yuan, 1999). This indicated 
that if ADAFs were obtained correctly, they would have been special cases 
of the advection solution.  

One importance that is often (wrongly) attributed to ADAF is that it 
(and only it, if press reports are to be believed) can distinguish 
between a black hole and a neutron star. The reason being that since
ADAF carries away dissipated energy along with it, the flow does not 
have to radiate. Thus the maximum luminosity of black holes should be 
less and lo and behold, it was so! One  (especially observers who are 
too happy to see some model fitting their observations) has to be 
extremely careful in this context. It is long known (Chakrabarti, 1990a) 
that black holes allow perfectly stable transonic flow solutions 
with constant specific energy but neutron stars do not. These have been 
found to be stable through numerical simulations (Chakrabarti \& Molteni,
1993; Molteni, Sponholz \& Chakrabarti, 1994). So it is no wonder that
the flow {\it can} enter through the horizon without radiating. In fact, 
this `non-radiating' aspect of ADAF was also pointed out by none other 
than Paczy\'nski (1998) who gave a `toy model' of ADAF nearly six years 
after ADAF was proposed only to highlight his earlier works of thick 
disks which were also non-radiating or nearly non-radiating. 

\section{Observational Challenges}

While identifying a black hole, one has to scrutinize the observational
results very carefully. There are several categories of observations. 

\noindent \underline{Category 1}: Some observation should be unique to the black 
holes only, or other objects should not give rise to the same observational 
results. For instance, it is well known that power-law hard tails have been
observed  in `black hole candidates' even in soft-states. The question 
is: how can this power-law be generated which extends up to hundreds of 
KeV when the electrons are cooler then a keV?. Theoretical answer is 
straight forward: since matter must enter into a black hole 
supersonically with the velocity of light, photons scattered by 
these relativistic electrons {\it must be energized} even if the 
electrons are cold.  This is called the `Bulk Motion Comptonization' 
or simply BMC. Chakrabarti \& Titarchuk (1995) showed that a power-law 
hard tail is produced in this process. Indeed, this power-lawhard tail was observed in
all the `originally' suspected black hole candidates as soon as they 
go to the soft state, i.e., in a situation when the electrons in 
CENBOL got cooler (Titarchuk et al. 1999). Incidentally, hot flows 
also produce energized photons through the so-called Thermal 
Comptonization (TC) process and creates a power-law slope. Thus, in 
the hard-state, both TC and BMC takes place, but power-law due to TC dominates. 

One could argue that the observational evidence that `black holes' 
are less luminous than neutron stars (Narayan, Garcia and McClintok, 1997)) as predicted by 
ADAF solution also identifies black holes. This argument is false.  On the 
one hand, in principle, outflows could also take away the inflowing 
matter even before it dives into the black hole and even before it had 
opportunity to radiate. This would also reduce the luminosity.  So the 
flow not necessarily sample the horizon and thus it is unsuitable 
to confirm a black hole's existence. Even if these ``observational results"
turn out to be correct, this fact alone does not say anything about whether 
an ADAF solution is correct or
applicable. It only means that the black holes do accrete through constant or nearly constant 
energy transonic flow solutions which are known for fifteen years.

\noindent \underline{Category 2}: In this category, the observations are interesting 
but because the horizon is not sampled, one cannot say anything about
whether one has seen a black hole or not. For instance, if the mass 
measurement is made through kinematics or through the measurement of 
the Doppler shift, and the compact core is found to be massive enough, 
it gives an indication that the primary could be a black hole. However, 
it is not a full proof.

\noindent \underline{Category 3}: There could be certain observations which could 
`baffle' the theoreticians after a source is `confirmed' 
to be a black hole (in the sense that it exhibited a Category 1 observation). 
Examples include the origin of quasi-periodic oscillations (QPOs) in black 
hole candidates. The power-density spectra of light curves clearly showed 
evidence of two types of flows: region with lower frequency and flat slope 
is from the flow farther away from the black hole, while the region with a 
higher frequency and steeper slope is from the flow closer to the black hole.  
The QPO occurs at or close to the break frequency.

The reason why the occurrences of QPOs is baffling is this: in neutron stars
QPOs were thought to be due to falling of lumps of matter from a Keplerian
disk onto the star surface and the QPO supposedly occurs at the beat frequencies.
There is no hard surface on a black hole. Why and how would QPO form? 
Only advective flow solution has the answer! It has been shown most 
convincingly that presence of cooling processes triggers oscillations 
of the CENBOL (Molteni, Sponholz and Chakrabarti, 1996; Chakrabarti, 
Acharyya and Molteni, 2004) which in turn processes variable high energy 
X-ray intensity after intercepting variable amounts of soft-photons from the pre-shock 
flow. The power density spectra of the computer simulated light curve show
exactly the same features (see, Chakrabarti, Acharyya and Molteni, 
this volume). This vindicates the assertions of an advective disk model.

\noindent \underline{Category 4}: There are observations which do not have direct 
relation to black holes. In fact, systems with neutron stars and black 
holes do exhibit them. But it would be the best if the accretion flow 
model, which explains Category 1-3 observations also explains this 
Category 4 observation to complete satisfaction. Case in point is the 
formation of jets and outflows. These flows are ubiquitous in any 
gravitating system. Some disk models, such as standard disks, which
have no preferred length scales will produce jets all over the place 
as in Blandford and Payne (1982) model. However, profuse jets are 
observed only when the spectrum is hard, i.e., when the Keplerian 
disk component is weak. Furthermore, the base of the jet has to 
be less than few tens of Schwarzschild radii (Junor, Biretta \& Livio, 1999). Such
puzzling situation is beautifully handled by the advective flow model: 
the CENBOL produces the jet and this is found to be present only in the 
hard states when the it is not cooled down by the profuse number of 
soft-photons (Chakrabarti 1999).  Also, the CENBOL is a few tens of 
Schwarzschild radii big. This explains the small base of the jets observed
in both the galactic and extragalactic black holes.

\section{Discussions and Conclusions}

We discussed in this review that different models present in the literature take care of 
only different isolated aspects of the black hole astrophysics. Models even vary from black hole
to black hole. A Keplerian disk gives a
soft X-ray bump (or a blue bump in the case of AGNs). A slim disk stabilizes the Keplerian disk
through radial motion at the inner edge. A thick disk produces a quasi-spherical region around 
a black hole which emits with a low efficiency except near the funnel wall. An ADAF could be 
valid in the limit of a vanishing accretion rate. However, an advective disk, which is 
a self-consistent solution of the governing equations, describe {\it all the aspects} of
the observations very accurately. The most exciting feature is the prediction of the
presence of the boundary layer
of a black hole, i.e., the CENBOL. It is produced behind the accretion shock and dissipates the
energy in the form of hard-radiation. A black hole remains in a hard-state if the CENBOL 
is hot, it goes to the soft-state if the CENBOL is cooled down. The inner edge of 
the CENBOL passes through the inner sonic point before disappearing into the hole. 
This bulk converging flow creates the hard power-law tail in black holes even in the 
soft state. The outflowing jet is produced only when the CENBOL is present. Thus,
this model naturally connects the properties of the outflows with the spectral properties.
The radial and vertical oscillation of the CENBOL produces Quasi-Periodic Oscillations
in black hole candidates. The breaks in the power density spectra are produced by the 
transition radius (where the Keplerian disk ends) and the shock radius (outer surface 
of the CENBOL). The oscillation frequencies of the single or occasional double shocks
become the frequencies of the QPOs. There are many other subtle issues which other models
do not even have scopes to discuss. For instance, when outflows form, the optical depth
of the CENBOL is reduced and spectral softening takes place (Chakrabarti 1998b). Similarly, when the
matter falls back to the CENBOL from failed outflows, the spectral hardening takes place.
These are also observed (Chakrabarti et al. 2002). Oscillation frequency is determined 
by the size of the CENBOL which also determines the size of the sonic radius of the outflow.
This, in turn, determines the time to raise the optical depth to $\tau\sim 1$ in the
outflowing  region. When the outflow below the sonic radius is optically thick, it cools down
by reprocessing the disk photons. This going into `on' and 'off' states in outflows, reflect into
burst-on and burst-off states of the light curves, especially in objects like GRS 1915+105
where the inflow and therefore outflow rates are high. 

One of the puzzling aspects of the observations from a black hole candidate is the 
presence of iron lines which appear to be red- and blue-shifted in accordance with a 
Keplerian velocity around a black hole. This cannot be a correct picture since
there are inconsistencies in relation to equivalent width and intensity. Chakrabarti
\& Titarchuk (1995) showed that these lines could be produced more easily if the
source is places in outflows rather than in disks. In future, more work needs
to be done to see if this is indeed the case. 

The advective disk solution does not comprise of a single component only. It is
usually a two component disk having the Keplerian component on the equatorial plane 
and the sub-Keplerian components sandwiching it. Of course the degree varies.
This TCAF model has been shown to be a successful solution for all the known black holes
(Smith et al, 2001, 2002). 

The fact that the advective disk solution should explain so many aspects of observations from 
black hole candidates should not surprise anyone. This is because it is a solution rather than a 
model. However, solutions have been obtained for simplified equation of state (polytropic) and
in most of the solutions magnetic field is left out completely. Generally, it is introduced
as a stochastic field, so that it contributes to the magnetic pressure and synchrotron radiation.
It is possible that large scale magnetic field is not required and/or does not play any significant role
in black hole physics. One unexplained aspect is the presence of very highly relativistic outflows 
(often the velocity reaching above 99.9\% of light) which is perhaps difficult to achieve pure hydrodynamically.
This does not imply that large scale poloidal magnetic field must be included into the advective disk
solution. Perhaps rapid collapse of toroidal flux tubes in quick successions can eject matter
quick enough to achieve such velocity. The so-called MTCAF (magnetic TCAF) has been proposed
(Nandi et al. 2002) to explain anti-correlation of X-ray and radio fluxes in black hole
candidates. Perhaps the same mechanism is responsible for acceleration and collimation (through the
whoop stress) of the jets.

\section*{Acknowledgments} 
We acknowledge partial supports from the DST project grant No.  SP/S2/K-15/2001 and a RESPOND project
of Indian Space Research Organization.

\end{document}